\documentclass[preprint]{elsarticle}
\usepackage{bm}% bold math
\usepackage[cp1251]{inputenc}
\usepackage[english]{babel}
\usepackage{amsfonts,amssymb,amsmath}
\usepackage{amsmath,graphicx}
\usepackage{graphicx,pifont} 
\usepackage{epsfig}
\title{Relativism versus statistics in the microscopic foundation of thermodynamics}
\author[ayz]{A.Yu.~Zakharov}% \corref{cor1} %\fnref{fn1}
\ead{Anatoly.Zakharov@novsu.ru}
\author[maz]{M.A.~Zakharov}
\ead{Maxim.Zakharov@novsu.ru}
\address{Yaroslav-the-Wise Novgorod State University, Veliky Novgorod, 173003, Russia}
\begin{document}
	\begin{abstract}

Highlights:
\begin{itemize}
	\item Relativistic effect of crystal dynamics ``freezing''.
	\item Non-statistical model of thermodynamic equilibration. 
\end{itemize}

The dynamics of oscillations of a one-dimensional atomic chain is investigated in the harmonic approximation, taking into account the relativistic effect of retardation of interactions. It is shown that all oscillations of the chain are damped. A non-statistical mechanism for the irreversible thermodynamic equilibration in many-body systems is proposed.

\end{abstract}
	
	\begin{keyword}
		Many-body systems dynamics; retarded interactions; irreversibility; thermodynamic equilibration 
		
		MSC classes: 80A05 \ 80A10 \ 37K60 \ 37L60 \ 70H40
		
		\PACS: 05.20.-y \sep 63.10.+a \sep 05.70.Ln \sep 45.50.Jf
	\end{keyword}
	
	\maketitle

\section{Introduction}

It is well known that classical thermodynamics is based on three main postulates (the first, the second, and the third principles of thermodynamics) and on two auxiliary postulates. There is no unified terminology regarding auxiliary postulates~\cite{Guggenheim,Uhlenbeck,Lieb1,Lieb2,Brown1}.

We will call ``The minus first postulate'' statement about the existence of a state of thermodynamic equilibrium of many-particle systems~\cite{Uhlenbeck}: ``An isolated mechanical system consisting of a very large number of molecules approaches thermal equilibrium, in which all "macroscopic" variables have reached steady values''. 
In the lectures of Uhlenbeck~\cite {Uhlenbeck} this statement is called ``The zeroth law of thermodynamics'', and in the paper~\cite{Brown1} it is called ``The minus first postulate''.

We will call ``The zeroth postulate of thermodynamics'' the statement about the transitivity of the equilibrium state of the systems: ``If system A is in equilibrium with the systems B and C separately then B and C are also in equilibrium between themselves''~\cite{Uhlenbeck}.

The creation of Gibbs' statistical mechanics~\cite {Gibbs} was aimed at microscopic foundation of thermo\-dynamics. Despite the outstanding advances of statistical mechanics in calculating and explaining the properties of matter, a number of problems in thermodynamics remain open.
\begin{enumerate}
	\item	First of all, this is the problem of foundation of the zeroth and the minus first principles of thermodynamics. In statistical mechanics, these principles are postulated in the same way as the application of the concept of probability to describe the evolution of dynamical systems in the absence of sources of stochastization~\cite{Kubo1}. 
	
	Statistical mechanics is based on the combination of the principles of classical mechanics and the concept of probability. There is a significant inconsistency between them, clearly formulated in the well-known paradoxes of the reversibility of Loschmidt and Zermelo. These paradoxes (in reality, these are the internal contradictions of statistical mechanics) have not yet been satisfactorily resolved~\cite{Strien}.
	In the works of Kac~\cite {Kac1, Kac2} a ring model was proposed as a counterexample to the probabilistic description of dynamical systems.
	
	\item There is a fundamental contradiction between the irreversible behavior of real many-body systems and the reversibility of classical dynamics.

	This fact raises serious doubts not only about the correctness of the combination of classical mechanics with the concept of probability, but also about any possibility of microscopic foundation of thermodynamics within the framework of classical mechanics.

	\item It should be noted that the irreversibility of the dynamics of a many-body system is a necessary, but generally speaking, not a sufficient condition for both the zeroth and the minus first postulates of thermodynamics. 
\end{enumerate}

In 1909, shortly after the appearance of statistical mechanics, the controversial work of Ritz and Einstein~\cite{Ritz} was published, the authors of which expressed mutually exclusive points of view on the nature of irreversibility: ``Ritz considers the restriction in the form of retarded potentials as one of the sources of the second law of thermodynamics, whereas Einstein assumes that irreversibility is based exclusively on the probabilistic foundations''~\footnote{``The Ritz-Einstein Agreement to Disagree''~\cite{Fritzius} }.

The aim of this work is to present a possible non-statistical mechanism of the thermodynamic equilibration process in many-body systems. To illustrate this mechanism, the dynamics of a one-dimensional chain of particles with retarded interactions between them is studied.

\section{Dynamics of a one-dimensional chain with retarded interactions between particles}

Consider a one-dimensional system of identical particles interacting with each other, whose stable equilibrium positions form an ideal lattice with the Born~--~von Karman boundary conditions~\cite{Born1}
\begin{equation}\label{x-n}
x_{n}^{\left( 0\right) } = n\, a\quad  \left(a = \mathrm{const},\quad  x_{n+N}^{\left( 0\right) } = x_{n}^{\left( 0\right) }, \quad  1 \leq n \leq N\right).
\end{equation}

The basic principles of the dynamic theory of crystal lattices with \textit{instantaneous interactions} between particles was developed mainly in the works of Born and his co-authors~\cite {Born1, Born2, Born3, Born4}. Further development of crystal dynamics was aimed at taking into account the features of crystal structures, models of interatomic potentials, defects in crystals, nonlinear effects, etc.~\cite{Brillouin,Maradudin,Kosevich}.

The local value of the potential of the field created by all particles at the site $ x_{n}^{\left(0 \right) } $ for instantaneous interactions, i.e. neglecting the retardation effect, has the form

\begin{equation}\label{phi-0}
\varphi\left( x_{n}^{\left(0 \right) } \right) = \sum_{\stackrel{n'}{\left( n'\not= n\right) }}   \, v\left(x_{n}^{\left(0 \right) }  - x_{n'}^{\left(0 \right) }\right). 
\end{equation} 
In this case, the dynamics of the system in the harmonic approximation is described by the equations
\begin{equation}\label{instant}
m \ddot{U}_{n}\left( t\right) = \sum_{n'>0} v'' \left( n' \right)\, \left[ U_{n-n'}\left( t\right)  -2 U_{n}\left( t\right)   +  U_{n+n'}\left( t\right)\right],  
\end{equation}
where $ {v}''\left( n\right) $ is the second derivative of the function $ v\left( x\right)  $  at $ x=n $,   $ U_{n} \left (t \right) $~is the displacement of the $ n $-th particle from its equilibrium position
\begin{equation}\label{U-n}
{U}_{n}\left( t\right) = x_{n}\left( t\right)  - x_{n}^{\left( 0\right) }\left( t\right), \quad \left| {U}_{n}\left( t\right) \right| \ll a. 
\end{equation}

It is known, that the solutions of the equations of crystal lattice dynamics with \textit{instantaneous interactions} in the harmonic approximation lead to the concept of phonons. The dispersion law of these quasiparticles depends on the crystal structure and interatomic potentials. 

However, real interactions between particles are not instantaneous. They always have a delay property due to the finite speed of propagation of interactions.
This property leads to a radical change in the dynamics even in the simplest case of a two-body problem, including the irreversible behavior of the system~\cite{Synge,Zakharov2019}.

To take into account the effect of retardation of interactions between particles of a one-dimensional lattice in the equation~\eqref{instant}, we perform the replacement
\begin{equation}\label{tau1}
U_{n\pm n'}\left( t\right) \longrightarrow U_{n-n'}\left( t - \tau\left(n'a  \right)  \right),
\end{equation}
where $ \tau \left(n'a \right) $  is the retardation time of interaction between points located at a distance of $ n'a $ from each other.

By virtue of the condition~\eqref{U-n}, we assume that the retardations of interactions between each pair of particles depend only on the equilibrium distances between them.
Since the retardations of interactions between points are proportional to the distance between them, we put
\begin{equation}\label{tau-n}
\tau\left(n'a  \right) = \frac{an'}{c}=\tau_{1}\,n',
\end{equation}
where $ c $  is the speed of propagation of interactions between particles, i.e. the speed of light, $ \tau_{1} $ is the delay time of the interaction between the nearest neighbors of the lattice.

Thus, the equations of the dynamics of a one-dimensional chain of interacting particles in the harmonic approximation, taking into account the retardation of interactions, has the form
\begin{equation}\label{1D-retard}
\left\lbrace 
\begin{array}{l}
{\displaystyle  m \ddot{U}_{n}\left( t\right) = \sum_{n'>0} v'' \left( n' \right)\, \left[ U_{n-n'}\left(  t\ - n' \tau_{1}  \right)  -2 U_{n}\left( t\right)   +  U_{n+n'}\left(  t\ - n' \tau_{1} \right) \right];  }\\
{\displaystyle U_{n+N}\left(t \right) = U_{n}\left(t \right) .}
\end{array}
\right. 
\end{equation}

\section{Solution of equations of dynamics for a chain with retarded interactions}
To obtain solutions of equations~\eqref{1D-retard}, we use the Euler substitution
\begin{equation}\label{1D-char}
U_{n}\left(t \right) = u_{n}\, e^{-i\omega t}
\end{equation}
and find the following system of equations
\begin{equation}\label{charact}
-m\omega^{2}u_{n} = \sum_{n'>0} v'' \left( n' \right)\left[  u_{n-n'}\, e^{i\omega\tau_{1}n'} - 2\, u_{n} +  u_{n+n'}\, e^{i\omega\tau_{1}n'} \right]. 
\end{equation}
We put now 
\begin{equation}\label{u-n}
u_{n} = u\, e^{ikan},
\end{equation}
and find
\begin{equation}\label{omega}
-m\omega^{2} = \sum_{n'>0} v'' \left( n' \right) \left[e^{-ikan'+i\omega\tau_{1}n'} -2 + e^{ikan'+i\omega\tau_{1}n'} \right]. 
\end{equation}
This \textbf{characteristic} equation with respect to $ \omega $ in the general case (that is, for $ \tau_{1} \neq 0 $) is a transcendental equation and the set of its roots is infinite. 

From the boundary condition~\eqref{x-n} it follows
\begin{equation}\label{k-N}
e^{ikaN} = 1,
\end{equation} 
therefore
\begin{equation}\label{kaN}
	k = 2\pi\frac{s}{aN}
\end{equation}
($ s $ is an arbitrary integer)
and
\begin{equation}\label{k}
- \frac{\pi}{a} \leq	k < \frac{\pi}{a}.
\end{equation}

Note that in the case when the retardation of interactions has a relativistic origin
\begin{equation}\label{tau}
\tau_{1} \sim a/c,
\end{equation}
the parameter
\begin{equation}\label{Omega-tau}
\varepsilon = \Omega_{0}\tau \sim v/c \ll 1
\end{equation} 
 is small ($ v $ is the characteristic speed of atomic motion). 
Therefore, from the infinite set of roots of the characteristic equation, a finite number of ``actual'' roots are distinguished.

As a result, the transcendental characteristic equation~\eqref{omega} is reduced to a quadratic equation with respect to the ``actual roots''
\begin{equation}\label{omega1}
\omega^{2}	+i \omega \frac{\tau}{m} \sum_{n'>0} n'\, v''\left( n'\right) \cos\left( kan' \right)  - \frac{1}{m} \sum_{n'>0}\, v''\left(n' \right) \left[1 - \cos\left(kan' \right)  \right] =0.
\end{equation}
In this approximation, the roots of the characteristic equation have the form
\begin{equation}\label{roots}
	\omega_{1,2}\left( k \right) = \pm \sqrt{\frac{2}{m} \sum_{n'>0}\, v''\left(n' \right) \sin^{2}\left( \frac{kan'}{2}\right)  } -i \frac{\tau}{2m} \sum_{n'>0}\, n' \, v''\left( n'\right) \, \cos\left( kan'\right). 
\end{equation}

Note that the imaginary part of these roots of the characteristic equation is negative, therefore the substitution of~\eqref {roots} in~\eqref {1D-char} leads to the conclusion that all the oscillatory degrees of freedom of the system are damped.

Compare the real and imaginary parts of $ \omega_{1,2}\left( k \right) $
\begin{equation}\label{Re-Im}
	\frac{\mathrm{Im}\ \omega\left( k\right) }{\mathrm{Re}\ \omega\left( k\right) }  = \frac{\displaystyle \frac{\tau}{2m} \sum_{n'>0} n'\, v''\left( n'\right) \cos\left( kan'\right) }{\displaystyle \sqrt{\frac{2}{m} \sum_{n'>0}\, v''\left(n' \right) \sin^{2}\left( \frac{kan'}{2}\right)  } }.
\end{equation}

The asymptotics of this ratio in a vicinity of the point $ k = 0 $ has the form
\begin{equation}\label{asymp}
	\frac{\mathrm{Im}\ \omega\left( k\right) }{\mathrm{Re}\ \omega\left( k\right) }	\approx \frac{\displaystyle  \sum_{n'>0} n'\, v''\left( n'\right) }{\displaystyle \sqrt{\sum_{n'>0}\, \left( n'\right)^{2} v''\left(n' \right)  } }\, \frac{\tau}{\sqrt{2m}\ \left| a\,k\right|} \to \infty.
\end{equation}
Therefore, the long-wave oscillations ($ \left| ak \right| \ll 1 $) damp most rapidly. When $ t \to \infty $, all oscillations in the chain stop.

\section{Discussion}

Retardation of interactions between particles leads to a catastrophic change in the dynamics of a one-dimensional crystal lattice, namely, to complete ``freezing'' of the motion of the particles in the system.
How can this seemingly paradoxical result be interpreted? Where does the kinetic energy of atoms go over time?

Within the framework of pre-relativistic physics, the essence of the potential energy of interactions between particles remained a kind of ``thing in itself''. It is some function that depends on the \textbf{instantaneous configuration} of the system and has an hidden origin~\footnote{
		In this regard, it is appropriate to note one of the first attempts to find a mechanical interpretation of the interaction of distant bodies: in the outstanding treatise~\cite{Hertz} Heinrich Hertz proved that the potential energy of interacting bodies is mathematically equivalent to the kinetic energy of hidden particles}.

The point is that, within the framework of the classical dynamics of a finite system with instantaneous interactions, there is no need to resort to the concept of a field to describe interactions between particles. Such a system admits a Hamiltonian description of the evolution of a system with a finite number of degrees of freedom.

Since the real interaction between atoms has a field nature, a complete description of the system of interacting atoms must contain both equations of particle dynamics and equations of field dynamics.
Because of the field, the complete set of degrees of freedom of a system containing even a finite number of particles is infinite.

In the papers~\cite{Zakharov2019a,Zakharov2020,Zakharov2020a}, the exact elimination of field variables in the equations of the dynamics of particles and the electromagnetic field was performed. As a result, the dynamics of a system of particles is described by a closed system of differential-functional equations of delayed type of non-Hamiltonian type.

Thus, the retardation of interactions between atoms is the mechanism by which the kinetic energy of particles is transferred to the energy of the field. The field through which the interaction between atoms takes place plays the role of an inexhaustible thermodynamic reservoir. This reservoir with an infinite number of degrees of freedom ensures the irreversible motion of the system of atoms to absolute rest, that is, to the ground state of the system.

The evolution of a system of atoms, as a subsystem of a complete system, consisting of atoms and a field, due to retarded interactions goes beyond the Hamiltonian theory. Therefore, strictly speaking, neither the Liouville theorem on the conservation of the phase volume, nor the Poincar\'{e} theorem on recurrences, nor the Liouville equation for the phase density, nor the theorem of uniqueness of the Cauchy problem have a place for a system of atoms.

Thus, the relativistic effect of the retardation of interactions between particles leads not only to the irreversibility of particle dynamics, but is also a mechanism that ensures the realization of ``the minus first postulate'' of thermodynamics, i.e. an irreversible process of  thermodynamic equilibration in both many-body and few-body systems.

\section*{Conclusion}

We are very grateful to Prof.~Ya.I.~ Granovsky, Prof.~V.V.~Uchaikin, and Dr.~V.V.~ Zubkov for the stimulating discussions and useful comments.

\end{document}